\documentclass[twocolumn,prc,aps,superscriptaddress,showpacs]{revtex4-1}
\usepackage{graphicx}
\usepackage{multirow}
\usepackage{amssymb}
\usepackage[utf8]{inputenc}
\usepackage[T1]{fontenc}
\usepackage{latexsym}
\usepackage{paralist}
\usepackage{float}
\usepackage{lipsum}

\begin{document}

\title{$\beta$-delayed three-proton decay of $^{31}$Ar}

\date{\today}

\author{A.A.~Lis}
\affiliation{Faculty of Physics, University of Warsaw, 02-093 Warszawa, Poland}

\author{C.~Mazzocchi}
\email{chiara.mazzocchi@fuw.edu.pl}
\affiliation{Faculty of Physics, University of Warsaw, 02-093 Warszawa, Poland}

\author{W.~Dominik}
\affiliation{Faculty of Physics, University of Warsaw, 02-093 Warszawa, Poland}

\author{Z.~Janas}
\affiliation{Faculty of Physics,  University of Warsaw, 02-093 Warszawa, Poland}

\author{M.~Pf\"utzner}
\affiliation{Faculty of Physics, University of Warsaw, 02-093 Warszawa, Poland}
\affiliation{GSI Helmholtzzentrum f\"{u}r Schwerionenforschung, 64291 Darmstadt, Germany}

\author{M.~Pomorski}
\affiliation{Faculty of Physics,  University of Warsaw, 02-093 Warszawa, Poland}

\author{L.~Acosta}
\affiliation{INFN, Laboratori Nazionali del Sud, Via S.~Sofia, Catania, Italy}
\affiliation{Instituto de F\'isica, Universidad Nacional Aut\'onoma de M\'exico, M\'exico D. F. 01000, Mexico.}

\author{S.~Baraeva}
\affiliation{Joint Institute for Nuclear Research, 141980 Dubna, Russia}

\author{E.~Casarejos}
\affiliation{University of Vigo, 36310, Vigo, Spain}

\author{J.~Du\'{e}nas-D\'{\i}az}
\affiliation{Department of Applied Physics, University of Huelva, 21071 Huelva, Spain}

\author{V.~Dunin}
\affiliation{Joint Institute for Nuclear Research, 141980 Dubna, Russia}

\author{J.~M.~Espino}
\affiliation{University of Seville, 41012 Seville, Spain}

\author{A.~Estrade}
\affiliation{University of Edinburgh, EH1 1HT Edinburgh, United Kingdom}

\author{F.~Farinon}
\affiliation{GSI Helmholtzzentrum f\"{u}r Schwerionenforschung, 64291 Darmstadt, Germany}

\author{A.~Fomichev}
\affiliation{Joint Institute for Nuclear Research, 141980 Dubna, Russia}

\author{H.~Geissel}
\affiliation{GSI Helmholtzzentrum  f\"{u}r Schwerionenforschung, 64291 Darmstadt, Germany}

\author{A.~Gorshkov}
\affiliation{Joint Institute for Nuclear Research, 141980 Dubna, Russia}

\author{G.~Kami\'nski}
\affiliation{Institute of Nuclear Physics PAN, 31-342 Kraków, Poland}
\affiliation{Joint Institute for Nuclear Research, 141980 Dubna, Moscow Region, Russia}

\author{O.~Kiselev}
\affiliation{GSI Helmholtzzentrum  f\"{u}r Schwerionenforschung, 64291 Darmstadt, Germany}

\author{R.~Kn\"obel}
\affiliation{GSI Helmholtzzentrum  f\"{u}r Schwerionenforschung, 64291 Darmstadt, Germany}

\author{S.~Krupko}
\affiliation{Joint Institute for Nuclear Research, 141980 Dubna, Russia}

\author{M.~Kuich}
\affiliation{Faculty of Physics, University of Warsaw, 02-093 Warszawa, Poland}
\affiliation{Faculty of Physics, Warsaw University of Technology, 00-662 Warszawa, Poland}

\author{Yu.~A.~Litvinov}
\affiliation{GSI Helmholtzzentrum  f\"{u}r Schwerionenforschung, 64291 Darmstadt, Germany}

\author{G.~Marquinez-Dur\'{a}n}
\affiliation{Department of Applied Physics, University of Huelva, 21071 Huelva, Spain}

\author{I.~Martel}
\affiliation{Department of Applied Physics, University of Huelva, 21071 Huelva, Spain}

\author{I.~Mukha}
\affiliation{GSI Helmholtzzentrum  f\"{u}r Schwerionenforschung, 64291 Darmstadt, Germany}

\author{C.~Nociforo}
\affiliation{GSI Helmholtzzentrum  f\"{u}r Schwerionenforschung, 64291 Darmstadt, Germany}

\author{A.K.~Ord\'{u}z}
\affiliation{Department of Applied Physics, University of Huelva, 21071 Huelva, Spain}

\author{S.~Pietri}
\affiliation{GSI Helmholtzzentrum  f\"{u}r Schwerionenforschung, 64291 Darmstadt, Germany}

\author{A.~Prochazka}
\affiliation{GSI Helmholtzzentrum  f\"{u}r Schwerionenforschung, 64291 Darmstadt, Germany}

\author{A.M.~S\'{a}nchez-Ben\'{\i}tez}
\affiliation{Department of Applied Physics, University of Huelva, 21071 Huelva, Spain}
\affiliation{Centro de F\'isica Nuclear da Universidade de Lisboa, 1649-003 Lisboa, Portugal}

\author{H.~Simon}
\affiliation{GSI Helmholtzzentrum  f\"{u}r Schwerionenforschung, 64291 Darmstadt, Germany}

\author{B.~Sitar}
\affiliation{Faculty of Mathematics and Physics, Comenius University, 84248 Bratislava, Slovakia}

\author{R.~Slepnev}
 \affiliation{Joint Institute for Nuclear Research, 141980 Dubna, Russia}

\author{M.~Stanoiu}
\affiliation{IFIN-HH, Post Office Box MG-6, Bucharest, Romania}

\author{P.~Strmen}
\affiliation{Faculty of Mathematics and Physics, Comenius University, 84248 Bratislava, Slovakia}

\author{I.~Szarka}
\affiliation{Faculty of Mathematics and Physics, Comenius University, 84248 Bratislava, Slovakia}

\author{M.~Takechi}
\affiliation{GSI Helmholtzzentrum  f\"{u}r Schwerionenforschung, 64291 Darmstadt, Germany}

\author{Y.~Tanaka}
\affiliation{GSI Helmholtzzentrum  f\"{u}r Schwerionenforschung, 64291 Darmstadt, Germany}
\affiliation{University of Tokyo, Japan}

\author{H.~Weick}
\affiliation{GSI Helmholtzzentrum  f\"{u}r Schwerionenforschung, 64291 Darmstadt, Germany}

\author{J.S.~Winfield}
\affiliation{GSI Helmholtzzentrum  f\"{u}r Schwerionenforschung, 64291 Darmstadt, Germany}

\begin{abstract}
The $\beta$ decay of $^{31}$Ar, produced by fragmentation of a $^{36}$Ar beam at
880 MeV/nucleon, was investigated. Identified ions of $^{31}$Ar were stopped in a
gaseous time projection chamber with optical readout allowing to record decay events
with emission of protons. In addition to $\beta$-delayed emission of one and two
protons we have clearly observed the $\beta$-delayed three-proton branch.
The branching ratio for this channel in $^{31}$Ar is found to be $0.07 \pm 0.02$\%.
\end{abstract}

\pacs{23.40.-s, 23.50.+Z, 27.30.+t}
%\keywords{Suggested keywords}
%Use showkeys class option if keyword
%display desire

\maketitle

\section{Introduction}
\label{sec:intro}

Nuclides at the proton drip-line are characterised by large $\beta^+$-decay $Q$ values which allow population of particle-unbound states in the daughter nuclei. This makes the process of $\beta$-delayed (multi) particle emission possible \cite{blank2008,RMP}. The probability of this phenomenon becomes particularly sizeable when the isobaric analogue state (IAS) is located above the respective threshold.
%Due to the large energy available and the low separation energy for (multi) particle emission, the %process of $\beta$-delayed emission of several nucleons from the daughter nucleus becomes possible.
%The study of $\beta$-delayed (multi-) proton decay has allowed an insight into the structure of such %exotic nuclei \cite{blank2008, borge1990, axelsson1998, axelsson1998a} since the identification of %beta-delayed proton ($\beta$p) from $^{25}$Si \cite{barton1963}, $\beta$-delayed two-proton ($\beta$2p) %from $^{22}$Al \cite{cable1983} and $\beta$-delayed three-proton ($\beta$3p) decay from $^{45}$Fe %\cite{miernik2007}.
The $\beta$-delayed single proton emission ($\beta p$), studied intensively over the last 50 years, has provided a wealth of information on the structure of neutron-deficient nuclei \cite{blank2008}.
The delayed emission of two protons ($\beta 2p$), discovered in 1983 \cite{cable1983}, now is known
to occur in 11 cases \cite{blank2008,mp1}. The interesting additional aspect of this decay mode is the
mechanism of two-proton emission which could proceed either sequentially or simultaneously.
Much less is known on delayed emission of more than two particles. Prior to the study
reported in this paper only two cases of $\beta$-delayed three-proton emission ($\beta 3p$)
were known: $^{45}$Fe \cite{miernik2007} and $^{43}$Cr \cite{pomorski2011}.
Both of them were observed with help of a very sensitive gaseous detector ---
the Optical Time Projection Chamber (OTPC) --- which records tracks of charged particles in space.
With this approach a single
good event is sufficient to establish a new decay channel. In this paper
we report the observation of the third case of $\beta 3p$ decay, $^{31}$Ar, accomplished
with this technique.

%since it was discovered almost 30 years ago \cite{langevin1986},
%providing a wealth of information on its %$\beta$2p decay \cite{bazin1992, borge1990, axelsson1998, %axelsson1998a, fynbo1999, fynbo2000}
%If several cases of $\beta$p and $\beta$2p emission are known throughout the proton-dripline, $\beta$3p %emission is still a very rare decay mode with only 3 cases identified to date: $^{45}$Fe %\cite{miernik2007}, $^{43}$Cr \cite{pomorski2011} and $^{31}$Ar \cite{pfutzner2012, koldste2014}.

The most neutron-deficient argon isotope known to date, $^{31}$Ar, was observed for the first
time almost 30 years ago \cite{langevin1986}. It decays with the half-life
of $T_{1/2} = 15.1(3)$~ms \cite{koldste2014}
and the energy $Q_{EC} = 18.3(2)$~MeV \cite{AME2012}. In such a large decay-energy window
many channels for delayed emission of protons are open. In fact, $^{31}$Ar is the most
extensively studied case of $\beta 2p$ decay \cite{borge1990, axelsson1998a, axelsson1998, fynbo2000, fynbo2002}. In addition, it was considered as a primary candidate for the $\beta 3p$ decay
and the search for this decay mode has an interesting history.
Already in 1992 Bazin et al. claimed the first observation of this channel in $^{31}$Ar \cite{bazin1992}.
In an experiment performed at GANIL Caen with $^{31}$Ar ions implanted into a segmented silicon
detector, the branching ratio for $\beta 3p$ channel was claimed to
be $\beta 3p = 2.1(10)$\% \cite{bazin1992}. However, a measurement made later at CERN-ISOLDE,
employing  a silicon detector array of larger granularity and efficiency, failed to observe
this decay mode. Only an upper limit for the $\beta 3p$ decay branch from the IAS state in $^{31}$Cl
to the ground state of $^{28}$Si was deduced to be $0.11$\% \cite{fynbo1999}.

Motivated by the success of applying the OTPC detector to studies of $\beta$-delayed
multiparticle emission \cite{miernik2007,pomorski2011}, we have used it to investigate
the delayed particles emitted in the decay of $^{31}$Ar. The preliminary results,
demonstrating clear and unambiguous evidence for the $\beta 3p$ events, were presented
in Ref. \cite{pfutzner2012}. Soon afterwards, the reanalysis of data collected previously
at CERN-ISOLDE fully supported this observation \cite{koldste2014,koldste2014b}.
Here we present results of the complete analysis of our experiment.

%Among these isotopes, $^{31}$Ar has been most extensively studied since it was discovered almost 30 years %ago \cite{langevin1986}, providing a wealth of information on its $\beta$2p decay \cite{bazin1992, %borge1990, axelsson1998, axelsson1998a, fynbo1999, fynbo2000} and discordant information on its $\beta$3p %decay \cite{bazin1992, fynbo1999, koldste2014}. In 1992 the first observation of $\beta$3p radioactivity %from $^{31}$Ar was announced with b($\beta$3p)=2.1(10)\% \cite{bazin1992}. This observation was not %confirmed in the subsequent measurements at ISOLDE, which led to b($\beta$3p)<1.1$\cdot 10^{-3}$ (99\% %C.L.) \cite{fynbo1999}. The first unambiguous identification of $\beta$3p decay branch from $^{31}$Ar %came in 2012 \cite{pfutzner2012} and it was immediately confirmed by Koldste et al. with a deduced %branching ratio b($\beta$3p)>=3.9(19)$\cdot 10^{-4}$ \cite{koldste2014}.
%In this paper we report on the full analysis of the preliminary data reported in \cite{pfutzner2012}, %with the determination of a new, improved value for the  $\beta$3p branching ratio (b$_{\beta3p}$).

\section{Experimental technique}
\label{sec:exp}
The experiment was performed at the GSI Helmholtzzentrum für Schwerionenforschung,
Darmstadt, Germany.
The $^{31}$Ar ions were produced in the fragmentation reaction of a $^{36}$Ar primary beam at 880 MeV/nucleon impinging on a 8~g/cm$^2$ beryllium target. The $^{31}$Ar fragments were separated by the Fragment Separator (FRS) \cite{hg1}.
The setting of the FRS was not the standard one.
Instead, the first half of the separator was used to select and form the secondary beam of $^{31}$Ar using a 5~g/cm$^2$ thick aluminum degrader located at the S1 focal plane which was shaped for achromatic focusing to the intermediate focal plane (S2). There this beam was impinging on a secondary target for a purpose of a different experiment which
will be reported elsewhere. The unreacted
ions of $^{31}$Ar were transported through the second half of the FRS to the final
focal plane (S4). There, after passing
through the standard FRS detectors, the ions were slowed down by a variable aluminum degrader,
and entered our detection system based of the gaseous Optical Time Projection Chamber (OTPC).
In front of the OTPC entrance window a $30 \times 30$~mm$^2$, 300 $\mu$m thick silicon detector was mounted and
an additional variable aluminum degrader to optimize the implantation of selected ions was installed.

The OTPC was developed at the University of Warsaw, primarily to study rare decay processes
with emission of charged particles, like 2p radioactivity \cite{km1,km2}. The unit used in the
present study was described in some details in Ref. \cite{mp1}. Here we give only the main
features and those details which are specific to the reported experiment.
The active volume of the chamber was filled with a gas mixture of~98~\%~Ar and 2~\%~N$_2$
at atmospheric pressure. The fiducial length of the detector was 31~cm.
A uniform, vertical electric drift field of 300 V/cm was maintained
inside the chamber. Primary ionization electrons, generated by incoming ions and their
charged decay products, were drifting towards the charge amplification stage made by a stack
of four GEM foils \cite{GEM} and a wire mesh anode. Before reaching the anode, the electrons stimulate
the gas atoms to emit light. This light was detected by a CCD camera and a photomultiplier tube (PMT).
The CCD image provided a projection of the particles tracks on the horizontal plane integrated over the exposure time. The PMT signal, sampled by a digital oscilloscope, gives the total intensity of light as a function of time. This provides information on the time sequence of recorded events. In addition, when
the drift velocity of electrons in the given field and gas mixture is known, the vertical
coordinate can be determined. Then, by combining data from the CCD and the PMT, the track of a particle can be reconstructed in three dimensions ~\cite{km1,mp1}.

The detector could operate in a low- or a high-sensitivity mode depending on the
potential applied to a special gating electrode mounted between the active volume and the
amplification stage. In the low-sensitivity mode the detector is protected against
large charges generated by heavy ions. Within 100~$\mu$s from the
triggering signal, the mode was switched to high sensitivity, optimal
for less ionising particles like protons \cite{mp1}. After the exposure
the detector was returned to the low-sensitivity mode.

Ions arriving to the OTPC were identified in-flight using the time-of-flight (TOF) and
the energy-loss ($\Delta E$) information. The TOF was measured by means of plastic scintillators mounted
at the S2 and the S4 focal planes of the FRS. Each scintillator was read
from the left (L) and the right (R) side with respect to the beam axis. Two signals were
formed by means of time-to-amplitude (TAC) converters, corresponding to the time
between left (LL) and right (RR) readouts. The energy loss ($\Delta E$) was measured
by the silicon detector in front of the OTPC. These three identification signals were
sampled by a digital oscilloscope and their waveforms were stored. Later, in the off-line
analysis the amplitudes of these signals were determined and the average of both TOF
values was taken as the final time-of-flight value for the ion.

In order to reduce the amount of data collected, a selective trigger based on the identification
information was applied. The trigger signal was generated only by those ions for which
the $\Delta$E and one of the TAC signals exceeded certain preset amplitudes. The identification plot
for ions which triggered the OTPC acquisition system is shown in Fig. 1a.
In all previous experiments with the OTPC system, the beam was turned off upon receiving the triggering
signal to protect the detector from other ions entering the active volume in the high sensitivity mode.
In this experiment such a protection was not possible due to the acceleration scheme of the SIS synchrotron.
The primary beam was impinging on the target in 1~s long spills slowly extracted from the SIS.
Thus, the reaction products passing through the FRS were coming in bunches correlated
with these spills. When the triggering ion arrived, all following ions from the same bunch
could enter the OTPC in the high sensitivity mode within the exposure time.
The identification signals for all such ions, in addition to the triggering ion, were recorded.
Fig.~1b shows the identification plot for all ions which entered the OTPC during
the total exposure time, including the triggering ions.

\begin{figure}[htbp]
%\vspace{0.5cm}
\includegraphics[width=0.45\textwidth]{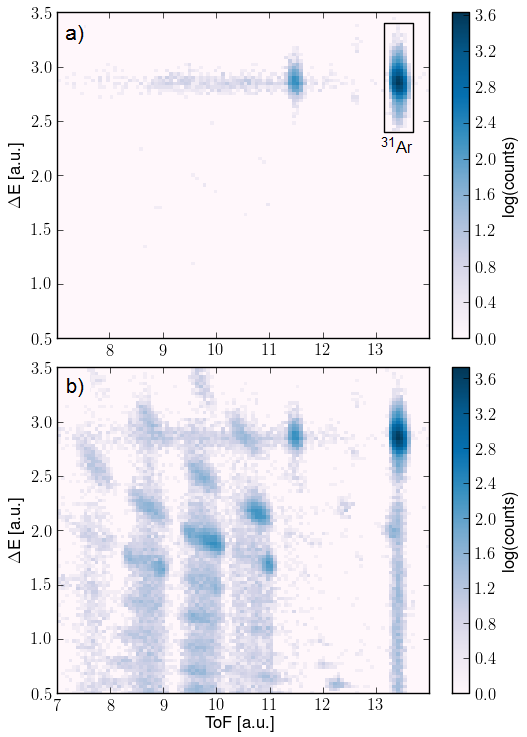}
\caption[id_plots] {The $\Delta$E--TOF identification plots for ions arriving to the
OTPC detector. a) Only the ions which triggered the data acquisition system.
The rectangular box shows the gate used to select the $^{31}$Ar events for further analysis.
b) All ions recorded by the OTPC during the experiment.}
\label{id_plots}
\end{figure}

Tracks of the ions which entered the OTPC after the trigger were "contaminating" the CCD image
and could obscure the decay events of the stopped $^{31}$Ar ions.
In order to remedy this problem and to reduce the number of particle tracks on CCD images,
we have applied a special acquisition mode by dividing the full CCD exposure
into five shorter frames. The CCD camera was working in the "movie"-like mode,
taking continuously subsequent images with a constant exposure time of $16$~ms. 
The dead-time between two consecutive frames due to read-out was 0.79 ms. 
When the trigger signal arrived, five consecutive frames,
starting with the one in which the triggering ion was present,
were validated and stored. After the entire exposure lasting 80~ms all
the collected data were written to a PC hard disk. The data for each event
consisted of five CCD images, the PMT waveform,
the three waveforms of the identification signals and the camera readout signal marking the
starting time of each CCD frame. An example set of five CCD images collected for one event,
picturing the $\beta 3p$ decay of $^{31}$Ar, is presented in Fig.~2.

\begin{figure*}[htb]
\includegraphics[width=\textwidth]{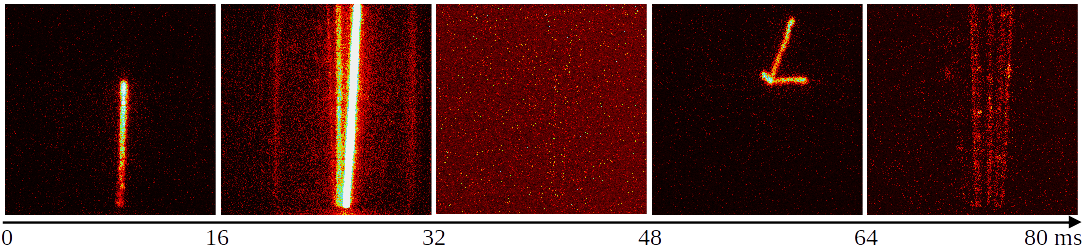}
\caption[frames] {(Color online) Example set of five consecutive CCD frames collected for one event. Each frame was exposed for 16~ms. In the first frame the track of an implanted $^{31}$Ar ion is visible.
In the second and in the
fifth frames tracks of contaminant ions passing through the chamber are seen,
while on the third frame nothing of interest happened.
In the fourth frame the decay of
the implanted ion by emission of three $\beta$-delayed protons is captured.}
\label{frames}
\end{figure*}

\section{Analysis and results}
\label{sec:results}
Events corresponding to the ions of $^{31}$Ar which triggered the OTPC acquisition were
selected by the software gate indicated by a rectangle in Fig.~1a.
The total number of such triggers during the data-taking time of five days was about 53000.
However, as a result of the high-energy fragmentation reaction and the large thickness
of all materials in the beam (target, degraders) the range straggling of the
reaction products was much larger than the effective thickness of the OTPC detector.
For the further analysis we have selected only those events in which the triggering $^{31}$Ar ion
was stopped inside the detector, between 10\% and 90\% of its length.
The latter limits were applied to make sure that all decays with emission of protons
will be clearly visible on the CCD image. In addition, only those events were taken into account
where on the first CCD frame no other ions than $^{31}$Ar were present.
This measure was taken to avoid any ambiguity in the assignment of the decay
event to the implanted ion. Finally, this procedure yielded about 21000 events
representing the proper implantation of a $^{31}$Ar ion.

\begin{figure}[htbp]
\includegraphics[width=0.5\textwidth]{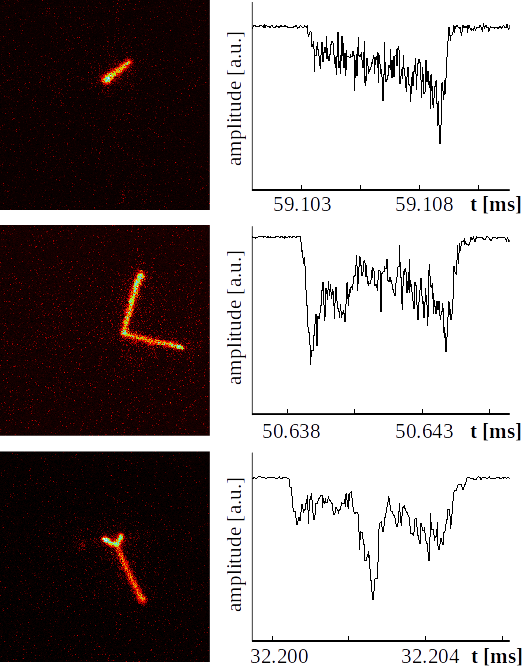}
\caption[event3p] {(Color online) Example events of $\beta$-delayed proton emission from $^{31}$Ar.
In the left column the CCD images of emitted particles are shown, while the corresponding
PMT waveforms are presented in the right column. Examples of $\beta p$, $\beta 2p$,
and $\beta 3p$ decays are shown in the top, the middle, and the bottom row, respectively. }
\label{event3p}
\end{figure}

Each event was inspected individually. The decay events picturing the emission
of a proton or a simultaneous emission of two and three protons were clearly observed.
Example events of these three categories are shown in Fig.~3. If the decay was
captured on a different CCD frame than the implantation, it was accepted only
when the coordinates of the decay vertex on the CCD image coincided with those
of the point of implantation. In addition, in case of emission of two and three
protons it was verified that the corresponding PMT waveform was consistent with the
scenario that all protons were emitted simultaneously, representing a single decay event.
The analysis yielded 13157 events of $\beta p$, 1729 events of $\beta 2p$, and 13 events
of $\beta 3p$. In the remaining events no decay signals were observed. In these cases either the decay occurred without emission of a proton, or the delayed proton(s) were emitted during the dead-time of the system or after the full exposure time. %Since $\beta$ electrons ionize too weakly to be detected by the OTPC and the
%total exposure time corresponded on average to 4.8 half-lives of $^{31}$Ar,
%we assume that in all these events a decay did occur but with no emission of
%delayed protons. 
Due to the finite measuring time and the dead-time between frames, the average probability to observe a decay in the five-frame time-window was 92.3\%. This value together with the number of decays allowed to determine the total branching ratios for all the decay channels observed. The results are collected in Table~\ref{table}.

\begin{table}
\caption{The total branching ratios for the observed decays of $^{31}$Ar.
         The given uncertainties are statistical.}
\begin{ruledtabular}
  \begin{tabular}{ccc}
		Channel & Events & Branching [\%] \\
		\hline
		$\beta 0p$ & 5984  & $22.6 (3)$$^a$ \\
        $\beta 1p$ & 13157 & $68.3 (3)$ \\
        $\beta 2p$ & 1729  & $9.0 (2)$ \\
        $\beta 3p$ & 13    & $0.07 (2)$ \\
  \end{tabular}
  \label{table}
  \end{ruledtabular}
\flushleft  $^a$ Value obtained by subtracting the sum of the $\beta 1p$, $\beta 2p$ and $\beta 3p$ branching ratios from 100\%.
\end{table}

The determined branching ratios for the $\beta 1p$ and $\beta 2p$ decays are in
agreement with the literature values of $62(2)$\% and $8.5(4)$\% \cite{bra}.
The uncertainties of our values are smaller which illustrates the advantage
of the OTPC detector for this kind of studies. Our results are obtained by a simple
counting of individual events which are unambiguously identified. No normalisations are involved.

The classification of 13 events as $\beta 3p$ decays was established with
high confidence. It might have happened, however, that some additional events of this
type were misclassified as $\beta 2p$ decays. This could happen if the topology
of the decay was particularly unfortunate, for example, if
one of the protons was emitted along the direction of the electric field and its
trace on the PMT waveform was obscured by signals from remaining protons.
Since it is difficult to estimate the probability of such scenario,
we give only the statistical uncertainty to the branching ratio for the
$\beta 3p$ decay. The value of this branching, however, should be considered
as a lower limit.

Our value for the $\beta 3p$ branching is consistent with the upper limit
of $0.11$\% established previously by Fynbo et al. \cite{fynbo1999}.
Recently, Koldste et al. \cite{koldste2014} reported the observation of the
$\beta 3p$ channel in $^{31}$Ar in the experiment performed at ISOLDE,
where the delayed particles were detected by means of a highly segmented array of
DSSSD detectors providing large granularity and efficiency \cite{Cube}.
The branching ratio for the $\beta 3p$ decay going through the IAS state
in $^{31}$Cl to the ground state of $^{28}$Si was determined to be $0.039(19)$\%.
In addition, it was estimated that this number represents roughly half
of all $3p$ emissions, while the rest of them goes through highly-lying
states above the IAS. Thus, the total branching observed in Ref. \cite{koldste2014}
amounts to about $0.08(4)$\%. Within error bars, this value agrees well
with our result.

In principle, from the data measured by the OTPC detector it is possible to reconstruct
the energy of an observed particle under condition, however, that the whole
track is contained within the fiducial volume \cite{mp1}. Unfortunately,
in the present experiment a large fraction of the delayed protons emitted in
the decay of $^{31}$Ar escaped the detector volume. Hence, here we report
only on the decay modes and their total branching ratios.

\section{Summary}
\label{sec:summary}
Beta decay of $^{31}$Ar was investigated at the GSI Fragment Separator by means of an optical-readout time-projection chamber. Decay channels with emission of delayed
protons were directly and clearly observed. The measured total branching ratios
for the $\beta p$ and $\beta 2p$ channels were found to be in agreement
with the literature data \cite{bra}. Thirteen events of the $\beta 3p$ emission
were unambiguously identified, yielding the branching ratio for this
decay mode of $0.07(2)$\%. This result is in good agreement with the value
reported recently by Koldste et al.~\cite{koldste2014}.

Although the probability of the $\beta 3p$ channel is small, it may have a large
impact on the $\beta$ decay strength distribution. This is because the
emission of three protons proceeds mainly from highly excited states
in the daughter nucleus, including levels above the isobaric analogue state.
As estimated in Ref.~\cite{koldste2014b}, the $\beta 3p$ transitions are
responsible for about 30\% of the total Gamow-Teller strength
distribution observed in $^{31}$Ar. For the exhaustive determination of
the $\beta$ decay strength, the complete information on all decay
modes, including all channels of delayed particle emission is mandatory.

The experimental technique of recording rare decay events with emission
of heavy charged particles by means of a gaseous OTPC detector, in combination
with the in-flight production of very exotic nuclei, proves to be
highly efficient and sensitive. Implantation of a single ion, identified
in flight, followed by an observation of its decay is in principle
sufficient to prove the occurrence of a decay channel. As a consequence
the branching ratios are measured with high accuracy. On the other
hand, determination of energy in the current version of the OTPC detector
is limited and cannot compete with arrays of silicon detectors.
Therefore, these two approaches to particle spectroscopy are complementary
and both have to be pursued.

%Future investigations of the energy- and angular-distributions of the
%delayed-protons emitted would require thicker gas mixtures to stop and study the higher energy %protons, while thinner gas mixtures would be needed to investigate the lower-energy end of the %spectrum. In the case of three or more particles emitted in the decay, unambiguous identification %can be achieved by replacing the optical readout with electronic one.

\section*{Acknowledgements}
The authors would like to thank the GSI staff for the high-quality beam delivered throughout the experiment. This work was partly supported by the Polish National Science Center under contract no. UMO-2011/01/B/ST2/01943. The participants from JINR are grateful for the grant RFBR No. 14-02-00090. A.A. Lis acknowledges support by the Polish Ministry of Science and Higher Education by the grant No. 0079/DIA/2014/43 ("Grant Diamentowy") and M. Pf\"utzner is grateful for a support from the Helmholtz International Center for FAIR (HIC for FAIR).

\end{document}